\documentclass[aps,prb,twocolumn,unsortedaddress,showpacs]{revtex4}%
\usepackage{graphicx}
\usepackage{epsfig}
\usepackage{amsmath}
\usepackage{amsfonts}
\usepackage{amssymb}

\newcommand{\bS}{\mbox{\boldmath$S$}}
\newcommand{\bQ}{\mbox{\boldmath\small$Q$}}

\newcommand{\bK}{\mbox{\boldmath\small$K$}}
\newcommand{\bk}{\mbox{\boldmath\small$k$}}
\newcommand{\bH}{\mbox{\boldmath$H$}}

\newcommand{\br}{\mbox{\boldmath$r$}}

\begin{document}

\title{
Modification of the classical Heisenberg helimagnet
by weak uniaxial anisotropy and magnetic field}

\author{I.~A.~Zaliznyak}
\affiliation{P.~L.~Kapitza Institute for Physical Problems, ul.~Kosygina 2,
117334 Moscow, Russia }

\author{M.~E.~Zhitomirsky}
\affiliation{L.~D.~ Landau Institute for Theoretical Physics, ul.~Kosygina 2,
117334 Moscow, Russia}
\affiliation{Institute for Solid State Physics, University of Tokyo, Tokyo 106
Japan}
\date{February 4, 1995}

\begin{abstract}

Classical ground state of the isotropic Heisenberg spin Hamiltonian on a
primitive Bravais lattice is known to be a single-$\bQ$ planar spin spiral. A
uniaxial anisotropy and external magnetic field distort this structure by
generating the higher-order Fourier harmonics at wave-vectors $n\bQ$ in the
spatial spin configuration. These features are not captured in the formalism
based on the Luttinger-Tisza theorem, where the classical ground state energy is
minimized under the ``weak'' condition on the length of the spins. We discuss
why the correct solution is lost in that approach and present an alternative
microscopic treatment of the problem. It allows to find the classical ground
state for general {\bQ} for both easy-axis and easy-plane uniaxial second-order
anisotropy, and for any orientation of the magnetic field, by treating the
effect of anisotropy (but not the field) as a perturbation to the exchange
structure. As a result, the classical ground state energy, the uniform
magnetization, and the magnetic Bragg peak intensities that are measured in
experiment, are calculated.

\end{abstract}


\maketitle

\section{Introduction}

For more than three decades, helical spin structures have been a subject of
intensive studies. Having been discovered theoretically in the pioneering works
of Yoshimori \cite{Yoshimori} and Villain \cite{Villain}, they were found
experimentally in a large variety of materials. Most extensive work was devoted
to the investigation and explanation of different ordered phases and phase
transitions in the rare-earth metals from Tb to Tm. These phases were shown to
result from the intricate distortion of the incommensurate, almost ferromagnetic
exchange helimagnet by temperature and strong crystal field anisotropies.
\cite{Nagamiya,Kaplan} However, for the reason discussed below, a classical
treatment, finally generalized by Lyons and Kaplan \cite{Lyons-Kaplan}, predicts
no such distortion in the ground state of the quasi-isotropic Heisenberg
Hamiltonian at $T=0$.

A simple but nontrivial example of the commensurate spiral is a triangular
magnetic ordering found in the hexagonal antiferromagnets of CsNiCl$_3$ type.
Their magnetic structure consists of six sublattices, three in each hexagonal
plane at an angle $\approx$120$^\circ$ to each other, whereas spins in the
adjacent planes are antiparallel. Such compounds have recently attracted much
attention because of their pronounced quasi-one-dimensional nature (exchange
between the adjacent spins along the hexagonal axis is much stronger than that
between the neighbors in the plane). Experiments
\cite{Yelon-Cox,Eibshutz,Zaliznyak} have demonstrated large deviations of the
staggered magnetization and susceptibility in these materials from the results
of the mean field (MF) calculations which should be valid for classical spins,
$S \gg 1$, and thus the importance of quantum fluctuations. Therefore, these are
also excellent examples of quantum helimagnets which also have been recently
studied in the literature. \cite{Chubukov*,Rastelli,Ohyama} Chubukov
\cite{Chubukov} and Tanaka \emph{et.~al.} \cite{Tanaka} calculated classical
spin reorientation process in magnetic field (see also
Ref.~\onlinecite{Abarzhi-Zaliznyak}) and the antiferromagnetic resonance spectra
in the hexagonal antiferromagnets of CsNiCl$_3$ type on the basis of
six-sublattice model. It appears that even for six sublattices the problem is
very difficult to handle in this way, and essentially no explicit analytical
results can be obtained for the spin-wave spectrum. Our motivation in the
present work was to develop a simpler and more general procedure for finding the
classical ground-state spin configurations, which would allow to perform
reasonably unsophisticated spin-wave calculations. We adopt the approach similar
to that used by Nagamiya \cite{Nagamiya} and Kaplan \cite{Kaplan}. Essentially,
it is extension of the iterative procedure first suggested by Cooper {\it et
al\/}. \cite{Cooper-Elliott} for the easy-plane, almost ferromagnetic spirals,
to a general helimagnet. However, keeping in mind its application to the
commensurate antiferromagnets, we will neither discuss here the
incommensurability effects, nor changes in the ordering wave-vector {\bQ} by the
anisotropy and magnetic field that occur for general {\bQ} position. These
problems, as well as the incommensurate-commensurate transitions, were treated
elsewhere in the framework of the phenomenological
approach.\cite{Dzyaloshinsky,Izyumov}

We start from the following spin Hamiltonian, which describes the
low-temperature magnetic properties of a large class of crystals where spins of
the magnetic ions are localized at lattice sites,
\begin{equation}
\label{hamiltonian}
\hat{\cal H} = \sum_{i,j} J_{ij}\,{\bf S}_i{\bf S}_j +
 D\sum_i(S_i^z)^2 - \gamma {\bf H}\sum_i{\bf S}_i\ .
\end{equation}
Here the first term is the Heisenberg exchange interaction, the second term
describes the lowest-order uniaxial anisotropy, and the last term is the Zeeman
energy ($\gamma$ is the gyromagnetic ratio). The anisotropy arises from the
magnetic spin-spin and spin-orbit interactions. Hence, the constant $D$
incorporates a small relativistic ratio $(v/c)^2$, where $v$ is the average
velocity of the electron in the atom, and $c$ is the velocity of light. This
constant is usually considered to be small in comparison with the exchange
coupling constants $J_{ij}$ which are of the electrostatic origin.  This
condition, however, can be violated in the quasi-low-dimensional magnetic
materials, where the superexchange path along some directions is indirect and
the corresponding constants $J_{ij}$ are very small. Although such cases, as
well as singlet ground state systems in which $D$ is intrinsically very large
may also be of interest, in the present paper we will always imply condition
$|D|\ll|J_{ij}|\neq 0$ fulfilled. We also restrict our attention to the cases in
which summation in (\ref{hamiltonian}) is performed over a single Bravais
lattice with $N$ sites.

In the spin-wave theory, the ground state and the excitation spectrum of quantum
Hamiltonian (\ref{hamiltonian}) are obtained in a semiclassical approximation,
based on $1/S$ expansion. The starting point for such calculation is to find an
equilibrium spin configuration with the minimum ground state (GS) energy
$E_{GS}$ in the classical limit, $S\rightarrow\infty$. In the case of the
isotropic Heisenberg Hamiltonian, a rather complete spin-wave theory describing
its low-temperature properties has been developed.
\cite{Chubukov*,Rastelli,Ohyama} However, the problem becomes extremely
complicated if the anisotropy which exists in real compounds is included
together with the magnetic field. In this case, even for classical helimagnets,
very little general results were actually obtained. Some particular cases were
successfully treated in Refs. \onlinecite{Nagamiya} and
\onlinecite{Cooper-Elliott}, but resulted in rather involved and hardly usable
expressions (in part, because an attempt to account for the six-fold anisotropy,
in addition to the easy-plane anisotropy, made problem much more complicated).
Here we present calculation of the ground state spin configurations for the
Hamiltonian (\ref{hamiltonian}) for different orientations of the magnetic
field, and taking into account the anisotropy (but not the field) as a
first-order perturbation of the exchange Hamiltonian.

\section{Failure of the weak condition}

It is known that ground state of a system of equivalent classical spins on a
simple Bravais lattice, in the exchange approximation, is a flat spin spiral
described by a wave-vector $\bQ$ (this includes ferromagnetism (${\bQ}=0$) and
antiferromagnetism (${\bQ}={\bK}/2$) as particular cases). A rigorous proof of
this result is obtained by solving the mathematical problem of finding the
absolute minimum of the function (\ref{hamiltonian}), which depends on $3N$
classical variables $S_i^\alpha$, under $N$ conditions,
\begin{equation}
\label{conditions} {\bS}_i^2 = S^2 \;, \forall i ,
\end{equation}
imposed on the lengths of the classical spins. This problem is solved by
introducing $N$ Lagrange multipliers $\lambda_i$. Switching to the Fourier
representation to use the lattice translational symmetry, one obtains the
following system of equations for spin configuration which minimizes the
Hamiltonian
\begin{eqnarray}
\label{minimum}%
NJ_{\bk}{\bS_{\bk}} + {\bf e}_z D S_{\bk}^z - \sum_{\bk'} \lambda_{\bk'}
{\bS_{\bk-\bk'}} = \frac{1}{2}\gamma {\bH} \delta_{{\bk},0}, \nonumber \\
\sum_{\bk'} {\bS_{\bk'} \bS_{\bk-\bk'}} = S^2 \delta_{{\bk},0} \ .
\end{eqnarray}
Here $J_{\bk}$, ${\bS_{\bk}}$, and $\lambda_{\bk}$ are the lattice Fourier
transforms of the functions $J_{ij}$, ${\bS}_i$, and $\lambda_i$, respectively;
${\bf e}_z$ is a unit vector along $z$ direction, and $\delta_{{\bk},0}$ is a
three-dimensional Kronekker symbol. From Eq. (\ref{minimum}) we easily obtain
the classical ground-state energy as
\begin{equation}
\label{g.s.energy}%
\frac{1}{N}\,E_{GS} = \lambda_0 S^2 - \frac{\gamma}{2}\,{\bH \bS}_0\ .
\end{equation}
There are two points to note here. First, Eq.~(\ref{minimum}) is a complicated
inhomogeneous system of nonlinear equations, for which no general solution is
easily found. Second, the resultant ground-state energy explicitly depends only
on the values of $\lambda_0$ and ${\bS}_0$ (on the latter only if ${\bH}\neq
0$), so all $\lambda_{\bk}$ for ${\bk}\neq 0$ seem to be irrelevant. This fact
encourages one to look for the solution for which the Lagrange multipliers have
the form $\lambda_{\bk} = \lambda_0\,\delta_{{\bk},0}$ and the system is largely
linearized. Evidently, such procedure is equivalent to taking into account only
one of the $N$ conditions given by the second relation in Eq. (\ref{minimum}),
or to replacing Eq. (\ref{conditions}) by the so-called ``weak'' condition,
$\sum_i{\bS}_i^2 = NS^2$. This standard approach was formalized by Lyons and
Kaplan \cite{Lyons-Kaplan}. Then, the solution is easily found in the form of a
single-{\bQ} spiral,
\begin{eqnarray}
\label{helix} {\bS_{\bk}} = {\bS}_0\delta_{{\bk},0} + {\bS_{\bQ}}\delta_{\bk,
\bQ} + {\bS^*_{\bQ}}\delta_{{\bk},-{\bQ}}\ , \nonumber \\
\lambda_0 =  N J_{\bQ} = N {\rm min}\{J_{\bk}\}\ ,
\end{eqnarray}
which is actually a correct result for $D\geq 0$ and {\bH} along the $z$ axis.
In this case one can easily choose vectors ${\bS}_0$ and ${\bS_{\bQ}}$ in
(\ref{helix}) that satisfy all $N$ ``strong'' conditions (\ref{conditions}) for
any $D$, or for any direction of the field, if $D=0$. Thus, according to the
Luttinger-Tisza theorem, the resultant spin configuration is the solution of
(\ref{minimum}), \emph{ie} it minimizes the energy (\ref{hamiltonian}) under the
``strong'' conditions (\ref{conditions}).

Unfortunately, except for the above important but simple cases, simple solution
obtained in this way appears to be physically meaningless. For example, for the
negative (easy-axis) anisotropy, the $H=0$ magnetic structure predicted with the
help of the ``weak'' condition is a collinear configuration with spins of
varying length, parallel to the $z$ axis.\cite{Kaplan} This prediction, in
particular, contradicts the experimental data in CsNiCl$_3$ and related
compounds, where the easy-axis anisotropy is known to align the spin plane
parallel to $z$-axis and to distort slightly the perfect 120$^\circ$ exchange
structure, initially described by a spiral with ${\bQ}=(\frac{1}{3},
\frac{1}{3}, 1)$, in reciprocal lattice units (rlu). In fact, the ``incorrect''
collinear phase, corresponding to the longitudinal helimagnet, also appears in
CsNiCl$_3$, in the narrow temperature interval between the two $T_N$ \cite{PhD}.

The reason for this contradiction and inconsistency of the trick with the
``weak'' condition is the following. It is easy to see, that by weakening the
conditions on the spin length one may obtain additional minima that do not
satisfy Eq. (\ref{conditions}). These artifacts are thrown away by verifying all
the conditions explicitly. However, some extrema of the initial problem
(\ref{minimum}) are also {\it lost} in this way. This can be visualized by
considering  the energy surface (\ref{hamiltonian}) in the space of spin
components, cut through by the ``physical'' manifold (\ref{conditions}). If the
real solution, \emph{ie} the global minimum, is one of the intersection points,
it will be inevitably missed by employing the ``weak'' condition, since it is
not an extremum of the function (\ref{hamiltonian}) in the extended space.

\section{Transformation to the local axes}

In order to perform the spin-wave calculation, one defines the local coordinate
axes, $x_iy_iz_i$, such that classical spin at a site $i$ points in $z_i$
direction. The spin operators are then transformed to these axes and expanded
into a series in Bose operators, \emph{eg} using Holstein-Primakoff
transformation. Because the local axis $z_i$ is a classical equilibrium
direction for spin at site $i$, the linear in spin deviations (\emph{ie}, in
operators $\hat{a}_i^+, \hat{a}_i$) terms are absent in the resultant boson
Hamiltonian. Starting from this point, one can develop a slightly different
procedure to find the classical ground state of the Hamiltonian
(\ref{hamiltonian}) and its subsequent decomposition into a series in Bose
operators. First, we transform the spins from the crystallografic $xyz$ frame to
the local coordinate axes $x_iy_iz_i$. In general, this is achieved by two
rotations: first, by an angle $\theta_i$ around the $y$ axis, and then by an
angle $\phi_i$ around the new $x'_i$ axis, as shown in Fig.~1. In the resultant
local frame the spin operators are
\begin{equation}
\label{transformation} \tilde{\bS}_i = \left\| {\bf T}_{2,i} \right\| \cdot
\left\|{\bf T}_{1,i}\right\| \cdot {\bS}_i = \left\|{\bf T}_{i}\right\| \cdot
{\bS}_i \ ,
\end{equation}
where the transformation matrices are defined by
\begin{eqnarray}
\label{matrices}
\left\|{\bf T}_{1,i}\right\| = \left\|\begin{array}{ccc}
\cos\theta_i & 0 & -\sin\theta_i\\ 0 & 1 & 0\\ \sin\theta_i & 0 &
\cos\theta_i \end{array} \right\| , \nonumber \\
\left\|{\bf T}_{2,i}\right\| = \left\|\begin{array}{ccc}
1 & 0 & 0\\ 0 & \cos\phi_i & -\sin\phi_i\\ 0 & \sin\phi_i &
\cos\phi_i \end{array}\right\| .
\end{eqnarray}
The transformed Hamiltonian becomes,
\begin{align}
\label{hamilt-tr}%
\hat{\cal H} = \sum_{i,j}J_{ij}\,\tilde{\bS}_i \left\| {\bf T}_{ij}^{({\rm
ex})}\right\| \tilde{\bS}_j + D\sum_i\tilde{\bS}_i\left\|{\bf T}_i^{(a)}\right\|
\tilde{\bS}_i \nonumber \\
- \gamma {\bH}\sum_i \left\| {\bf T}_{i} \right\|^{-1} \tilde{\bS}_i \ .
\end{align}
Here the matrices $\left\|{\bf T}_i^{(a)}\right\|$ and $\left\| {\bf
T}_{ij}^{({\rm ex})}\right\|$, which define the bilinear in spin components
forms coming from the anisotropy and the exchange coupling, respectively, are
given by
\begin{widetext}
\begin{equation}
\left\|{\bf T}_{ij}^{(a)}\right\| = \left\| \begin{array}{ccc} \sin^2\theta_i &
\sin\phi_i\sin\theta_i\cos\theta_i &
- \cos\phi_i\sin\theta_i\cos\theta_i\\
\sin\phi_i\sin\theta_i\cos\theta_i  & \sin^2\phi_i\cos^2\theta_i &
-\sin\phi_i\cos\phi_i\cos^2\theta_i \\
- \cos\phi_i\sin\theta_i\cos\theta_i &
-\sin\phi_i\cos\phi_i\cos^2\theta_i & \cos^2\phi_i\cos^2\theta_i
\end{array} \right\| ,
\end{equation}

and,

\begin{equation}
\left\|{\bf T}_{ij}^{({\rm ex})}\right\| = \left\|\begin{array}{ccc}
\cos\theta_{ij} & \sin\theta_{ij}\sin\phi_j & - \sin\theta_{ij}
\cos\phi_j \\
-\sin\theta_{ij}\sin\phi_i & \cos\phi_i\cos\phi_j + \cos\theta_{ij}
\sin\phi_i\sin\phi_j & \cos\phi_i\sin\phi_j - \cos\theta_{ij}
\sin\phi_i\cos\phi_j \\
\sin\theta_{ij}\cos\phi_i & \sin\phi_i\cos\phi_j - \cos\theta_{ij}
\cos\phi_i\sin\phi_j & \sin\phi_i\sin\phi_j + \cos\theta_{ij}
\cos\phi_i\cos\phi_j \end{array} \right\| ,
\end{equation}
\end{widetext}
where $\theta_{ij} = \theta_i-\theta_j$. The angles $\theta_i$ and $\phi_i$ must
be chosen so as to cancel the terms $\tilde{S}^x_i \tilde{S}^z_j$ and
$\tilde{S}^y_i\tilde{S}^z_j$ which give rise to the linear, in $\hat{a}_i^+,
\hat{a}_i$, contribution to the Hamiltonian (\ref{hamilt-tr}). This leads to the
following system of $2N$ equations:
\begin{widetext}
\begin{eqnarray}
 & & \sum_i 2 J_{ij}\cos\phi_i\sin\theta_{ij} =
D\cos\phi_j\sin 2\theta_j + \tilde{\gamma} H_x\cos\theta_j -
\tilde{\gamma} H_z\sin\theta_j \ , \nonumber  \\
 & & \sum_i 2 J_{ij}\left( \cos\phi_j\sin\phi_i -
\sin\phi_j\cos\phi_i\cos\theta_{ij} \right) \nonumber \\
 & & \mbox{} = D\sin 2\phi_j\cos^2\theta_j - \tilde{\gamma}
H_x\sin\phi_j\sin\theta_j + \tilde{\gamma} H_y\cos\phi_j -
\tilde{\gamma} H_z\sin\phi_j\cos\theta_j \ ,
\label{eq-angles}
\end{eqnarray}
\end{widetext}
where $\tilde{\gamma} = \gamma/$S. Evidently, this system for the angles
$\theta_i$ and $\phi_i$ cannot be solved for arbitrary $D$ and $H$.  As it was
mentioned above, only the cases $D=0$ and $D>0$, ${\bH}\parallel z$ can be
treated explicitly \cite{Yoshimori,Ohyama}. However, we can start with the
Heisenberg exchange Hamiltonian in a magnetic field as a zero approximation and
then look for the corrections, expanding them in powers of $D$. A very similar
approach was developed by Andreev and Marchenko \cite{Andreev-Marchenko} in the
framework of the phenomenological Lagrangian theory. They found a rather simple
and general solution of the problem in the leading order of perturbation.

%***************************Figure 1*****************************************
\begin{figure}[t]
\vspace{0.in}
\begin{center}
\label{examples_of_modulated_J}%
\includegraphics[width=2.5in]%,height=6.4in
{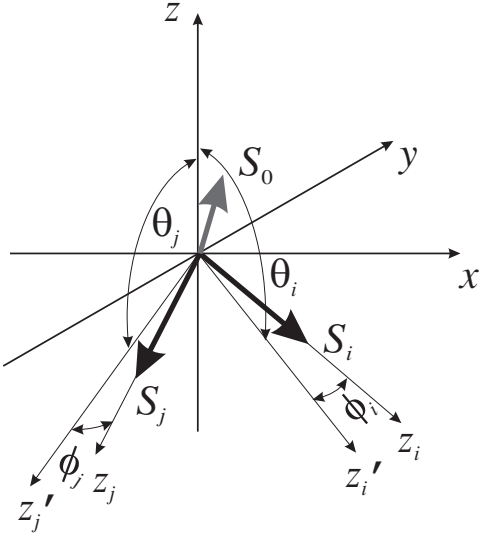}%
\caption{Transformation to the local axes for the distorted exchange spiral. }
\end{center}
\vspace{-.2in}
\end{figure}
%***************************Figure 1*****************************************

\section{Easy-axis anisotropy ($D<0$)}

First, we consider the more complicated case of the easy-axis anisotropy, which
leads to the most general distortion of the exchange structure. Without the loss
of generality we can set the magnetic field to lie in the $yz$ plane. In this
case the ``right'' unperturbed spiral should contain the $z$ axis already at
$H=0$. Therefore, we look for the solution of Eq. (\ref{eq-angles}) in the form,
\begin{equation}
\label{ea-angles}
\left\{ \begin{array}{lcr}
\theta_i & = & {\bQ \br}_i + \delta\theta_i \\
\phi_i & = & \phi_0 + \delta\phi_i
\end{array} \right. , \ \ \ \ \
\sin\phi_0 = \frac{\tilde{\gamma}H}{2(J_0 - J_{\bQ})} \ .
\end{equation}
In most cases we will consider only the leading (\emph{ie}, the lowest order in
$D$) corrections $\delta\theta_i$, $\delta\phi_i$. Thus, substituting
(\ref{ea-angles}) into (\ref{eq-angles}), it is sufficient to expand the
trigonometric functions to the first order in $\delta\theta_i$, $\delta\phi_i$.
As a result, we obtain the following equations for the deviations
\begin{widetext}
\begin{eqnarray}
 & & \sum_i 2J_{ij} (\cos\phi_0\cos({\bQ \br}_{ij})
\delta\theta_i -\sin\phi_0\sin({\bQ \br}_{ij})\delta\phi_i)
\nonumber \\
 & & \ \ \ \mbox{} = 2J_{\bQ}\cos\phi_0\delta\theta_j +
D\cos\phi_0\sin (2{\bQ \br}_j) - \tilde{\gamma} H_z(\cos({\bQ \br}_j)
\delta\theta_j + \sin({\bQ \br}_j)) \ ,
\nonumber  \\[2mm]
 & & \sum_i 2J_{ij}[(\cos^2\phi_0 + \sin^2\phi_0
\cos({\bQ \br}_{ij}))\delta\phi_i + \sin\phi_0
\cos\phi_0\sin({\bQ \br}_{ij}) \delta\theta_i] \nonumber \\
 & & \ \ \ \mbox{} = 2(J_0\sin^2\phi_0 + J_{\bQ} \cos^2\phi_0)
\delta\phi_j + D\sin\phi_0\cos\phi_0 (1+\cos (2{\bQ \br}_j))
\nonumber \\
 & & \ \ \ \ \mbox{}- (\tilde{\gamma} H_y\sin\phi_0 + \tilde{\gamma}
H_z\cos\phi_0\cos({\bQ \br}_j)) \delta\phi_j + \tilde{\gamma}
H_z\sin\phi_0\sin({\bQ \br}_j) \delta\theta_j
\label{eq-deltas}%
\end{eqnarray}
\end{widetext}
which are rather easy to solve. Indeed, upon convolution with $J_{ij} \cos ({\bQ
\br}_{ij})$, the variations $\delta\theta_i$ and $\delta\phi_i$ should yield the
trigonometric functions $\cos ({\bQ\br}_j)$, $\sin ({\bQ \br}_j)$, $\cos (2{\bQ
\br}_j)$, $\sin (2{\bQ \br}_j)$, \emph{etc}, which must cancel the right-hand
side of the Eq. (\ref{eq-angles}). Thus, they should be searched in the form of
an expansion, $\delta\theta_i = \sum_n (\alpha_n \cos (n{\bQ \br}_i) + \beta_n
\sin (n{\bQ \br}_i))$, where the order of the coefficients $\alpha_n$, $\beta_n$
is $H^n$ and $|D|^{n/2}$.

An important remark is due here.  One can also look for the solution of Eq.
(\ref{eq-angles}) with $\theta_i = \psi + {\bQ \br}_i + \delta\theta_i$, instead
of Eq. (\ref{ea-angles}). It differs by rotation of the arbitrarily chosen
``first'' spin, and thus of the whole spin structure, by an angle $\psi$ within
the plane of the spiral. Evidently, in the leading approximation (to the first
order in $D$) all such structures have the same energy, as it was the case for
the pure exchange. This fact reflects the remaining continuous degeneracy of the
ground state with respect to such rotations. Moreover, by performing the
corresponding expansions, it can be shown in a straightforward algebraic way
that this degeneracy exists for all orders in $H$ and $\sqrt{|D|}$ up to $n$
which satisfies the condition $n{\bQ}\equiv 0$, if such $n$ exists. This
condition means that spin structure is a commensurate spiral and can be
described in terms of $n$ sublattices. In this case the continuous degeneracy
corresponding to spin rotations within the spiral plane is lifted in the order
$H^n$ and $|D|^{n/2}$. The ground state preserves only $n$-fold degeneracy
corresponding to the arbitrary choice of the ``first'' spin, \emph{ie}, to the
rotation of all spins in the exchange structure by an angle $(2\pi m)/n$ in the
plane of the spiral. In the case of the incommensurate helimagnet, where this
discrete symmetry is absent, the continuous degeneracy of the ground state is
not lifted.

The above conclusions can be established from very general arguments. Evidently,
a perturbation of the $n$-sublattice exchange spiral by adding
$D\sum_i\cos^2(\psi+\theta_i)$ and $H\sum_i\cos(\psi+\theta_i)$ to the exchange
Hamiltonian is invariant with respect to the transformation
$\psi\rightarrow\psi+(2\pi m)/n$. Thus, the perturbed ground-state energy should
not contain harmonics below $\cos (n\psi)$ and $\sin (n\psi)$. Hence, explicit
dependence on the angle $\psi$ in its Taylor expansion with respect to the
perturbation parameters $D$ and $H$ only appears in the $n/2$-th and $n$-th
order terms, respectively. Another way to establish this fact is based on
exchange symmetry arguments \cite{Andreev-Marchenko}. Exchange structure with
$n$ sublattices has an $n$-fold rotation axis perpendicular to the spin plane.
Thus, the lowest-order perturbation of the exchange energy which has appropriate
symmetry and which lifts such rotations must be proportional to $\cos (n\psi)$.
In the case of the uniaxial anisotropy which already contains the two-fold
symmetry, it is achieved in the $n/2$-th order of perturbation. Magnetic field,
on the other hand, can lift the $n$-fold rotation axis characteristic of
exchange spiral only in the $n$-th order of perturbation. With the continuous GS
degeneracy removed, the energy gap appears in the corresponding (formerly)
Goldstone mode. This gap is induced by an ``effective field'' $\sim |D|^{n/2}$,
or $\sim H^n$, and, therefore, is of the order $\sim |D|^{n/4}$, and $\sim
H^{n/2}$, respectively.

This situation was closely studied in the hexagonal CsNiCl$_3$-type
antiferromagnets, which are our reference in the present paper. It was shown,
that in the easy-axis case the structure with $\psi = 0$, described by Eqs.
(\ref{ea-angles}), is stabilized by the anisotropy in the third order of
perturbation, and that the magnon gap $\sim|D|^{3/2}$ appears.
\cite{I.A.Zaliznyak,CsMnI} For the easy-plane antiferromagnetic spiral in the
transverse field the situation is different (as we discuss later), and the
structure with $\psi=\pi/2$ is realized. \cite{Chubukov,Abarzhi-Zaliznyak}
Below, we consider the solution (\ref{ea-angles}), which is stabilized by the
easy-axis anisotropy, $D<0$, for two important orientations of the magnetic
field.

\subsection{Magnetic field perpendicular to $z$-axis}

This field orientation gives rise to the most general distortion of the exchange
spiral. Deviations of the angles satisfying (\ref{eq-deltas}) are found in the
form
\begin{align}
\label{ea-perp}%
\delta\theta_i = \vartheta \sin (2{\bQ \br}_i) , \; \delta\phi_i =
\frac{\displaystyle D} {\displaystyle 2(J_0 - J_{\bQ})}\,{\rm tg}\phi_0 +
\varphi \cos (2{\bQ \br}_i) , \nonumber \\
\vartheta = \frac{\displaystyle D}{\displaystyle J_{3 \bQ} - J_{\bQ}} +
\frac{\displaystyle D}{\displaystyle J_{2\bQ} - J_{\bQ}} {\rm tg}^2\phi_0 , \;
\varphi = \frac{\displaystyle D}{\displaystyle J_{2\bQ} - J_{\bQ}} {\rm
tg}\phi_0 .
\end{align}
Note, that according to these expressions, the anisotropy causes a small
modulation of the spiral winding angle, $({\bQ \br}_i)$, which propagates as
$2({\bQ \br}_i)$. This naturally results in appearance of the harmonics at
wave-vectors $\pm 2{\bQ}$ and $\pm 3{\bQ}$ in the lattice Fourier transform of
the ground-state spin arrangement, in agreement with the early results of
Nagamiya {\it et\ al\/}, \cite{Nagamiya} and Cooper and Elliott.
\cite{Cooper-Elliott} Another point to be mentioned here is that small
variations (\ref{ea-perp}), which are treated here in the linear approximation,
actually contain terms which increase with the magnetic field (the most dramatic
is the second term in $\vartheta$, proportional to $H^2$). Therefore, although
these expressions are valid up to very high fields, they fail near the spin-flip
transition, at
\begin{equation*}
\tilde{\gamma}H_s \sim 2( J_0 - J_{\bQ})\left( 1 + \frac{D}{2(J_{2\bQ} -
J_{\bQ})}\right) .
\end{equation*}
In fact, this failure is indicative of a phase transition to a new spin
arrangement of the ``fan'' type, that was considered in detail by Nagamiya
\emph{et al}. \cite{Nagamiya} A simple way to study this transition and the
behavior of the system in the vicinity of the spin-flip field $H_s$, is to
expand the angles $\phi_i$ in small deviations  from $\pi /2$. This approach was
applied in Ref.~\onlinecite{Abarzhi-Zaliznyak}, where the field-induced
transitions in the easy-plane triangular antiferromagnet CsMnBr$_3$ were
studied.

The classical ground-state energy corresponding to the spin arrangement of the
global minimum found above can be easily calculated from Eqs.
(\ref{matrices})--(10),
\begin{align}
\label{Egs1}%
E_{GS} & = NS^2\left(J_{\bQ} + \frac{1}{2}\, D - (J_0 - J_{\bQ} +
\frac{1}{2}\,D)\sin^2\phi_0 \right) \nonumber \\
& = NS^2\left(J_{\bQ} + \frac{1}{2}\,D\right) - \frac{1}{2}\, N
\chi_{\perp}(\gamma H)^2 \ ,
\end{align}
where we have introduced the transverse spin susceptibility per site, as
follows,
\begin{align}
\label{chi-perp}%
\chi_{\perp} & = \frac{1}{2(J_0 - J_{\bQ}) - D} \nonumber \\
& = \frac{1}{2(J_0 - J_{\bQ})} \left(1 + \frac{ D}{2(J_0 - J_{\bQ})}\right) +
O\left( D^2\right) .
\end{align}
Spin structures are usually studied in neutron scattering experiments, by
measuring the intensities of the magnetic Bragg peaks. At each $\bk$, these are
proportional to the structure factor, \emph{ie} the square of the absolute value
of the corresponding Fourier transform of spin density, ${\bS_{\bk}}$, which,
therefore, can be called a Bragg amplitude. Thus, for comparison with the
experiment, it is important to calculate the nonzero Bragg amplitudes
${\bS_{\bk}}$ which result from the solution (\ref{ea-perp}). Introducing the
unit vectors ${\bf e}_x$ and ${\bf e}_y$ along the $x$ and $y$ axes,
respectively, we obtain,
\begin{align}
{\bS}_0 & = {\bf e}_y S\sin\phi_0 \left(1 + \frac{D}{2(J_0 -
J_{\bQ})}\right) = \chi_{\perp}(\gamma{\bH})\ , \nonumber \\
{\bS_{\bQ}} & = \frac{1}{2} ({\bf e}_z - i{\bf e}_x) S \cos\phi_0 \left(1 -
\frac{D}{2(J_0 - J_{\bQ})}{\rm tg}^2\phi_0
\right) \nonumber  \\
- ({\bf e}_z & + i{\bf e}_x) S \cos\phi_0 \left( \frac{D}{4(J_{3\bQ} - J_{\bQ})}
+ \frac{D}{2(J_{2\bQ} -
J_{\bQ})}{\rm tg}^2\phi_0\right) , \nonumber \\
{\bS}_{2\bQ} & = {\bf e}_y S\sin\phi_0\frac{D}
{2(J_{2\bQ} - J_{\bQ})} \ , \nonumber \\
{\bS}_{3\bQ} & = ({\bf e}_z - i{\bf e}_x) S\cos\phi_0\, \frac{D}{4
(J_{3\bQ}-J_{\bQ})}\ . \label{Sk-ea-perp}
\end{align}
In calculating the Bragg intensities, $I_{\bk}\sim |{\bS_{\bk}}|^2$, one has to
keep in mind that vectors ${\bS_{\bQ}}$ and ${\bS}_{3\bQ}$ have a nonzero
imaginary part.

\subsection{Magnetic field along $z$-axis; low fields}

This case seems to be less complicated, since the distortion of the exchange
spiral (\ref{helix}) occurs only at low fields, $H < H_c \sim \sqrt{|D|J}$, and
is such that it leaves the spins co-planar.  This means that $\phi_i = 0$, and
that the second equation in (\ref{eq-angles}) and in (\ref{eq-deltas}) is
automatically satisfied. The first equation is reduced to a rather simple
relation, from which, to the leading order in $D$ and $H$, the small deviations
$\delta \theta_i$ are,
\begin{align}
\label{ea-par1}%
\delta\theta_i & = \vartheta\sin (2{\bQ \br}_i) + \eta\sin ({\bQ \br}_i) ,
\nonumber \\
\vartheta & = \frac{D}{J_{3\bQ} - J_{\bQ}}\; , \;\;\eta = -
\frac{\tilde{\gamma}H}{J_0 + J_{2\bQ} - 2J_{\bQ}} .
\end{align}
It is clear from these expressions, that a leading order is insufficient for our
approximation. To be consistent to the first order in the anisotropy constant
$D$, one has to perform the expansions to the second order in $\eta \sim H\leq
\sqrt{|D|J}$. Such improved procedure leads to the following corrected
expression for $\vartheta$,
\begin{equation}
\label{ea-par2} \vartheta = \frac{D}{J_{3\bQ} - J_{\bQ}} + \eta^2 \left(
\frac{J_{2\bQ} - J_{\bQ}}{J_{3\bQ} - J_{\bQ}} - \frac{1}{4} \right) .
\end{equation}
The corresponding ground-state energy and the parallel susceptibility per spin
are
\begin{align}
\label{Egs,chi-par}%
E_{GS} & = NS^2\left(J_{\bQ} + \frac{1}{2}\,D - \frac{(\tilde{\gamma}H)^2}{4(J_0
+ J_{2\bQ} - 2J_{\bQ})} \right) , \nonumber \\
\chi_{\parallel} & = \frac{1}{2(J_0 + J_{2\bQ} - 2J_{\bQ})} ,
\end{align}
so that we can rewrite, $\eta = - 2\chi_\parallel (\tilde{\gamma}H)$. The Bragg
amplitudes are also easy to calculate,
\begin{align}
\label{Sk-ea-par1}%
{\bS}_0 = & \; {\bf e}_z S\,\frac{\tilde{\gamma} H} {2 (J_0 + J_{2\bQ} -
2J_{\bQ})} = \chi_\parallel \gamma{\bH} , \nonumber\\
{\bS_{\bQ}} = & \frac{1}{2}({\bf e}_z - i{\bf e}_x) S \left[1 -
\frac{D+\eta^2(J_{2\bQ} - J_{\bQ})} {2(J_{3\bQ} - J_{\bQ})}\right] \nonumber \\
+ & \frac{i}{2}{\bf e}_x S\left[\frac{1}{2}\eta^2 - \frac{D+\eta^2(J_{2\bQ} -
J_{\bQ})}{J_{3\bQ}-J_{\bQ}} \right] , \nonumber\\
{\bS}_{2\bQ} = & - ({\bf e}_z - i{\bf e}_x) S\,
\frac{\tilde{\gamma}H}{4(J_0+J_{2\bQ}-2J_{\bQ})} \nonumber \\
= & - \frac{1}{2}\chi_\parallel\gamma H({\bf e}_z - i{\bf e}_x) , \nonumber \\
{\bS}_{3\bQ} = & ({\bf e}_z - i{\bf e}_x) S\,
\frac{D+\eta^2(J_{2\bQ}-J_{\bQ})}{4(J_{3\bQ}-J_{\bQ})}\ .
\end{align}
Actually, we can calculate the homogeneous component of the spin density,
$\bS_0$, (\emph{ie} the magnetization) with better accuracy than in Eq.
(\ref{Sk-ea-par1}), to $\sim |D|^{3/2}$. To be completely consistent in this
order, one should add to $\delta\theta_i$ terms like $\beta\sin (3{\bQ \br}_i)$,
with $\beta \sim |D|^{3/2}$. However, retained in the first order, these terms
would not contribute to ${\bS}_0$. For our purpose it is therefore sufficient to
take $\delta\theta_i$ in the form (\ref{ea-par1}), and simply retain more terms
in the subsequent expansions. In this way we obtain the following correction to
${\bS}_0$,
\begin{align}
\label{nonlinear_S0}%
{\bS}_0 = & \chi_\parallel \gamma{\bH} \left[ 1 - \chi_\parallel D \left( 3 -
4\frac{J_{2\bQ} - J_{\bQ}}{J_{3\bQ} - J_{\bQ}} \right)\right. \nonumber \\
+ & \left. (4\tilde{\gamma} H)^2\chi_\parallel^3 ( J_{2\bQ} - J_{\bQ})
\left(\frac{J_{2\bQ} - J_{\bQ}}{ J_{3\bQ} - J_{\bQ}} - \frac{1}{2} \right)
\right].
\end{align}
This expression includes the leading nonlinearity of the magnetization, which is
often important for comparison with experiment.

\subsection{Magnetic field along $z$-axis; high fields}

This is the simplest and the best-known case. At some critical field, $H_c \sim
\sqrt{|D|J}$, a usual spin-flop transition takes place, and structure becomes
the same as that of the magnetized exchange spiral, (\ref{helix}),
\begin{align}
\label{ea-helix}%
{\bS_{\bk}} = &\; {\bf e}_z S\sin\phi_0 \delta_{{\bk},0} + ({\bf e}_x + i{\bf
e}_y)S\cos\phi_0\delta_{\bk, \bQ} \nonumber \\
+ &\; ({\bf e}_x - i{\bf e}_y)S\cos\phi_0\delta_{\bk, -\bQ} \ .
\end{align}
In this case,
\begin{align}
\label{Egs-par1}%
E_{GS} = & NS^2\left(J_{\bQ} - \frac{(\tilde{\gamma}H)^2}{4(J_0 - J_{\bQ} + D)}
\right) , \nonumber \\
\sin\phi_0 = & \frac{\tilde{\gamma}H}{2( J_0 - J_{\bQ} + D)} .
\end{align}
These expressions hold up to a full saturation. The corresponding spin-flip
field, $H_s$, is defined from the condition $\sin\phi_0 = 1$. Comparing the
expressions (\ref{Egs,chi-par}) and (\ref{Egs-par1}) for the ground state
energy, one obtains the following universal formula for the spin-flop field
$H_c$,
\begin{equation}
\label{Hc} \tilde{\gamma}H_c = \sqrt{2|D| (J_0 - J_{\bQ})\left( 1 + \frac{J_0 -
J_{\bQ}}{J_{2\bQ} - J_{\bQ}} \right)} .
\end{equation}

\section{Easy-plane anisotropy ($D>0$)}

For the easy-plane Heisenberg helimagnet, spin reorientation in magnetic field
reveals no new features in addition to those described in the previous section.
All spin configurations occurring in this case have their exact analogs for the
easy-axis helimagnet. Therefore, it is sufficient to establish the
correspondence between the cases. The simplest case is that of magnetic field
directed along the $z$ axis. The spin arrangement in this case appears to be
exactly the same as in Sec.~4.3, and is described by the same expressions, Eqs.
(\ref{ea-helix}) and (\ref{Egs-par1}). However, here we must keep in mind that
constant $D$ is now positive.

If the field is applied in the easy plane, \emph{eg} along the $x$ axis, the
spin arrangement in low fields is similar to that described in Sec.~4.2. This
correspondence, however, is not as direct as the one above. Since all spins lie
in the same plane, the solution of Eqs. (11) is obtained with $\theta_i = \pi/2$
and $\phi_i = {\bQ \br}_i + \delta\phi_i$. The first equation is then
automatically satisfied, while the second one, for $\phi_i$, becomes exactly the
same as equation for $\theta_i$ in the case of a small field along the $z$-axis
in the easy-axis case, $D<0$, $H\parallel z$, $H<H_c$, but with $D=0$. Since now
the symmetry breaking in the (easy) spin plane is different, we must consider
two possible solutions.

{\bf A.} First, we assume $\delta \phi_i = \eta\sin {\bQ \br}_i + \vartheta\sin
{2\bQ \br}_i$, which is exactly the same as considered in Sec.~4.2. In this case
we find that $\eta$ and $\vartheta$ are given by Eqs. (18) and (19), but with
$D=0$. It is clear from the transformation Eq. (7) that the resultant Fourier
components of the spin density are given by the same expressions (21), only with
${\bf e}_x$ and ${\bf e}_z$ replaced by ${\bf e}_y$ and ${\bf e}_x$,
respectively. Evidently, such solution can be stabilized by the magnetic field
in the case of a nearly ferromagnetic helimagnet, where spins tend to align
along the field direction.

{\bf B.} Now consider the solution with $\delta \phi_i = \pi/2 + \eta\cos {\bQ
\br}_i - \vartheta\sin 2{\bQ \br}_i$. Parameters $\eta$ and $\vartheta$ are
again given by Eqs. (18) and (19).  As discussed above, in our approximation
this solution is degenerate with the previous ({\bf A}), and differs from it
only by the phase multipliers in Bragg amplitudes,
\begin{widetext}
\begin{eqnarray}
\label{Sk-ep}%
{\bS}_0 & = & \chi_\parallel\gamma{\bH} \left[1+(4\tilde{\gamma}H)^2
\chi_\parallel^3 (J_{2\bQ}-J_{\bQ}) \left(\frac{J_{2\bQ}-J_{\bQ}} {J_{3\bQ} -
J_{\bQ}} - \frac{1}{2} \right)\right] , \nonumber \\
{\bS_{\bQ}} & = & \frac{1}{2} ({\bf e}_y + i{\bf e}_x) S\left[ 1 - \eta^2
\frac{J_{2\bQ} - J_{\bQ}}{2(J_{3\bQ} - J_{\bQ})} \right] - \frac{1}{2} {\bf e}_y
S\eta^2 \left[ \frac{1}{2}- \frac{J_{2\bQ} - J_{\bQ}}{J_{3\bQ} - J_{\bQ}}
\right] , \nonumber \\
{\bS}_{2\bQ} & = & ({\bf e}_x - i{\bf e}_y) S \frac{\tilde{\gamma} H}{4(J_0 +
J_{2\bQ} - 2J_{\bQ})} = \frac{1}{2}\chi_\parallel\gamma H({\bf e}_x-i{\bf e}_y)
, \nonumber \\
{\bS}_{3\bQ} & = & - ({\bf e}_y + i{\bf e}_x)S\eta^2\, \frac{J_{2\bQ} -
J_{\bQ}}{4(J_{3\bQ}-J_{\bQ})} .
\end{eqnarray}
\end{widetext}
It was shown by Chubukov \cite{Chubukov} for the case of a six-sublattice
commensurate helimagnet CsMnBr$_3$, that it is a type-{\bf B} structure that is
stabilized by magnetic field in the sixth order, $\sim H^6$. A non-zero
frequency, corresponding to the magnon gap $\sim H^3$, appears in the magnetic
resonance spectrum. Magnetic field favors this solution for an antiferromagnetic
spiral because in this case it tends to align all spins perpendicular to its
direction. In a general case of $n$-sublattice antiferromagnetic spiral,
structure (\ref{Sk-ep}) is stabilized by the magnetic field in the $n$-th order,
as we have discussed in Sec.~4.

Above the spin-flop transition which occurs for ${\bH}\perp z$ and $D>0$ at same
$H_c$, given by (\ref{Hc}), the spin structure becomes exactly the same as that
discussed in Sec.~4.1 (assuming ${\bH}\parallel y$), and is described by the
same expressions, but with $D>0$.

\section{Conclusion}

In this paper we presented a general pertutbative treatment of the spin
reorientation process for a weakly anisotropic spin spiral described by the
classical Heisenberg spin Hamiltonian (1) in a magnetic field. The calculations
were performed up to the first order in the anisotropy constant $D$. Previously,
each particular case of this general problem was treated individually, typically
in the framework of the multi-sublattice model. Many examples of such approach
are found in the literature devoted to calculating the spin reorientation in
``triangular'' antiferromagnets on hexagonal lattice.
\cite{Chubukov,Tanaka,Abarzhi-Zaliznyak} Apart from the loss of generality, in
many important non-trivial cases such calculations are very difficult and lead
to involved formulas which depend on a particular setup of the exchange
couplings. As a result, in most cases neither the spin-waves spectrum, nor even
the magnetic resonance frequencies, can be calculated in an explicit analytical
form on the basis of the sublattice model. In contrast, the description
presented here is quite general, and the results are rather simple. They provide
a solid basis for spin-wave calculations and for the estimates of the
spin-fluctuation contribution to the ground-state energy which can often be very
important (even for selecting the correct ground state). Although here we
calculated the spin configurations only for two principal orientations of the
magnetic field, the derivation can be easily generalized for any field
orientation. To this end, one must simply retain all relevant terms in Eqs.
(11), (13), and admix the $\sin ({\bQ \br}_i)$ and $\cos ({\bQ \br}_i)$
rotations to the angle variations $\delta\phi_i$, $\delta\theta_i$. Thus, for
example, some results reported in Ref.~\onlinecite{Abarzhi-Zaliznyak} can be
obtained in a more general way.

As discussed above, one can treat the magnetic structures and the
long-wavelength spin dynamics within the perturbation approach based on exchange
approximation in the framework of the phenomenological macroscopic Lagrangian
theory, as was proposed by Andreev and Marchenko. \cite{Andreev-Marchenko} For
the systems described by the Hamiltonian (1) the leading relativistic
corrections are equivalent to our treatment of the anisotropy. The macroscopic
theory has the advantage of being less sensitive to the fluctuation effects (the
mean-field ground state is chosen better, to partially account for the
fluctuations). However, it also has two substantial disadvantages in comparison
with the microscopic description presented here. First, it is not aimed at (and
is not suitable for) calculating a microscopic spin-density distribution, or a
spin-wave spectrum in the whole Brillouin zone. Therefore, the results cannot be
directly compared with neutron scattering experiments which are the most
sensitive and powerful tool for studying spin systems. Second, to reproduce the
higher-order effects, such as nonlinear magnetization given by Eq.
(\ref{nonlinear_S0}), macroscopic theory requires introducing additional
phenomenological constants. It is clear, that microscopic theory which is
developed here in the classical-spin approximation is far from being complete,
and requires additional treatment of quantum corrections. This point, as well as
calculation of the spin-wave spectra, are beyond the scope of this paper, and
will be subjects of our further studies.

\bigskip
\begin{acknowledgments}

We wish to thank L.-P.~Regnault, J.~Villain, and A.~F.~Andreev for interesting
discussions of the problem and for illuminating remarks. We also thank
A.~S.~Borovik-Romanov, L.~A.~Prozorova, A.~I.~Smirnov, and S.~S.~Sosin for
reading the manuscript and for useful comments. The work of I.~A.~Z. was
supported by Grant M3K000 of the International Science Foundation. M.~E.~Zh.
acknowledges financial support from the Japan Society for the Promotion of
Science.

\end{acknowledgments}

\end{document}